\documentclass[a4paper,12pt]{article}
\usepackage{amssymb}
\usepackage{graphicx}
\usepackage{amsmath}
\usepackage{indentfirst}
\usepackage{mathrsfs}

\begin{document}

\title{Path-Integral Bosonization of Massive Gauged Thirring Model}
\author{R. Bufalo$^1$\thanks{%
rbufalo@ift.unesp.br} and B.M. Pimentel$^1$\thanks{%
pimentel@ift.unesp.br}~ \\
%EndAName
\textit{{$^{1}${\small Instituto de F\'{\i}sica Te\'orica (IFT/UNESP), UNESP
- S\~ao Paulo State University}}} \\
\textit{\small Rua Dr. Bento Teobaldo Ferraz 271, Bloco II Barra
Funda, CEP 01140-070,S\~ao Paulo, SP, Brazil}}
\date{}
\maketitle

\begin{abstract}
In this work the bosonization of two-dimensional massive gauged Thirring model in the path-integral framework is presented. After completing the
bosonization prescription, by the fermionic mass expansion, we perform an analysis of the strong coupling regime of the model through
the transition amplitude, regarding the intention of extending the previous result about the isomorphisms, at quantum level, of the massless
gauged Thirring model to the massive case.
\end{abstract}

\section{Introduction}

Bosonization of $(1+1)$-dimensional field models has been widely investigated along the years \cite{5}; and has its historical roots
in the investigations of Klaiber on Thirring Model \cite{1}, and Lowestein and Swieca on Schwinger model \cite{2}. And, since Coleman's
proof of the equivalence between the massive Thirring and sine-Gordon models \cite{6}, the concept of bosonization has been claimed
as a remarkable tool to obtain some nonperturbative information of $(1+1)$-dimensional theories; the bosonization
prescription is now well-understood and rigorously established at both operator and functional frameworks \cite{7,20}.
The bosonization of two-dimensional models at finite temperature also receives largely attention nowdays \cite{16}.

Although it was a common belief, that the bosonization was a exclusive property of $(1+1)$-dimensional theories, where there is,
indeed, no spin to distinguish fermions from boson; the need of a powerful tool to extract nonpertubative content
from $(2+1)$-dimensional field theories, leads to a development of the bosonization to such theories \cite{21}; which
has its intrinsic pathologies and also, rich and beauty properties. However, the most bosonization treatments in
$(2+1)$-dimensional theories regard to a perturbative calculation of the fermionic determinant. Those $(2+1)$-dimensional theories,
abelian and non-abelian ones, had a renew interest at different perspectives which had lead to a growing in its development in the last few years \cite{8}.

At both scenarios, $(1+1)$-dimensional and $(2+1)$-dimensional theories, lie into: the nonperturbative or pertubative evaluation of the fermionic determinant and the noninvariance of the measure under chiral transformations \cite{10}, the crucial role in the path-integral bosonization. It is the chiral Jacobian which conceives the Wess-Zumino term in the abelian and non-abelian $(1+1)$-dimensional field theories \cite{20}. These features of the path-integral prescription will be explored, here, to perform the bosonization of a massive Thirring-like model which has local gauge invariance; such model is known as massive gauged Thirring model (MGTM).

The gauged Thirring model (GTM) was originally proposed as a generalization of usual Thirring model, where the local gauge
symmetry was implemented by the Hidden Local Symmetry technique \cite{11,15}. However, one of the most interesting features of GTM was pointed out by
Kondo in his proposal of GTM \cite{12}; where not only an auxiliary scalar field is introduced but a kinetic term for the gauge field as well.
The GTM, at classical level and strong coupling regime, behaves as electrodynamics with fermions (for $g\rightarrow\infty$) or,
as a fermionic current-current self-interaction (for $e^{2}\rightarrow\infty$) field theory; where, at $(1+1)$-dim., the first model
is know as Schwinger model (SM), and the second, as Thirring model (TM). Furthermore, it was recently proved that the equivalence between
GTM and SM and TM, at $(1+1)$-dim., it is also present at quantum level \cite{14}. Such
equivalence was proved through of a nonpertubative quantization, and further analysis of respective Green's functions and Ward-Fradkin-Takahashi identities.

Therefore, inspired by those points stressed above, we will perform here the path-integral bosonization of the $(1+1)$-dim. MGTM.
After defining the dynamics of the MGTM by its Lagrangian density, we study and analyze some important points as: how the gauge field
decouples from the fermionic ones, the choice of an appropriated gauge choice, in particular, how the $R_{\xi}$-gauge
is crucial for the results; the mass expansion in the transition amplitude which allows not only, the separation on the
boson and fermion sectors -- which simplifies the calculation of the vacuum expectation value of operators, through Wightman's functions --
but also, afterwards, establishes and performs the analysis by the strong coupling limits, of the isomorphisms present in the MGTM.
Our main reason for studying the bosonization of MGTM lies in extending the previous result of isomorphism between the massless models and also to
the massive ones. We shall perform, in the Sect.~\ref{sec:1}, the derivation of the bosonization of massive gauged Thirring model in the path-integral framework and prove, in the Sect.~\ref{sec:2}, its equivalence with massive Schwinger and Thirring models. In the Sect.~\ref{sec:3} we present our final remarks and perspectives.

\section{Path-Integral Bosonization}
\label{sec:1}

The dynamics of MGTM is defined by the following Lagrangian density \cite{12,14},
\begin{equation}
\mathcal{L}=\bar{\psi}\left( i\widehat{\partial }+\widehat{A}\right) \psi -m%
\bar{\psi}\psi +\frac{1}{2g}\left( A_{\mu }-\partial _{\mu }\theta \right)
^{2}-\frac{1}{4e^{2}}F_{\mu \nu }F^{\mu \nu },  \label{eq 1}
\end{equation}%
which is invariant under the local gauge transformations,%
\begin{equation}
\psi ^{\prime }\left( x\right) =e^{i\sigma \left( x\right) }\psi \left(
x\right) ,~A_{\mu }^{\prime }\left( x\right) =A_{\mu }\left( x\right)
+\partial _{\mu }\sigma \left( x\right) ,~\theta ^{\prime }\left( x\right)
=\theta \left( x\right) +\sigma \left( x\right) .  \label{eq 1.1}
\end{equation}%
The $\theta $-field into Eq.$\left( \ref{eq 1}\right) $ was introduced
following the St\"{u}ckelberg procedure to incorporate the local gauge
symmetry. Also, the system is defined into a Minkowskian $(1+1)$-dimensional
space-time; with the following choice of representation for the Dirac $%
\gamma $-matrices and also the metric:%
\begin{equation}
\gamma _{0}=\sigma _{1}=\left(
\begin{array}{cc}
0 & 1 \\
1 & 0%
\end{array}%
\right) ;\gamma _{1}=-i\sigma _{2}=\left(
\begin{array}{cc}
0 & -1 \\
1 & 0%
\end{array}%
\right) ;\gamma _{5}=\gamma _{0}\gamma _{1}=\sigma _{3}=\left(
\begin{array}{cc}
1 & 0 \\
0 & -1%
\end{array}%
\right) ;\eta _{\mu \nu }=\left(
\begin{array}{cc}
1 & 0 \\
0 & -1%
\end{array}%
\right) ,
\end{equation}%
the $\gamma $-matrices satisfy the algebra:
\begin{equation}
\{\gamma _{\mu },\gamma _{\nu }\}=2\eta _{\mu \nu };\gamma _{\mu }\gamma
_{5}=\epsilon _{\mu \nu }\gamma ^{\nu };\epsilon _{\mu \nu }=\left(
\begin{array}{cc}
0 & 1 \\
-1 & 0%
\end{array}%
\right) .
\end{equation}%
It is possible to write the gauge field, following the Helmholtz's theorem, as%
\begin{equation}
A_{\mu }\left( x\right) =\partial _{\mu }\eta \left( x\right) -\tilde{%
\partial}_{\mu }\phi \left( x\right) ,  \label{eq 1.2}
\end{equation}%
where we have defined the notation $\tilde{v}_{\alpha }=\epsilon _{\alpha \mu }v^{\mu }$. A
suitable gauge condition for the MGTM is the $R_{\xi }-$gauge which has the form:
\begin{equation}
R_{\xi }=\partial _{\mu }A^{\mu }\left( x\right) +\frac{\xi }{g}\theta
\left( x\right) ;  \label{eq 1.3}
\end{equation}%
an important feature, inherent in the choice of the $R_{\xi }-$gauge, is that in
the quantization procedure of MGTM, it is allowed the $\theta $-field to decouple
from the other fields. Indeed, the gauge condition $\left( \ref{eq 1.3}%
\right) $, combined with $\left( \ref{eq 1.2}\right) $,\ allow us write:%
\begin{equation}
\square \eta \left( x\right) +\frac{\xi }{g}\theta \left( x\right) =0.
\label{eq 1.4}
\end{equation}%
Now, with the decomposition of gauge field $\left( \ref{eq 1.2}\right) $, we
have an interaction between the fermionic fields with the scalar ones,
$\eta $ and $\phi $; thus, in the way to cancel out, partially, such coupling between the fields,
we perform the following change in the fermionic variables,%
\begin{equation}
\psi \left( x\right) =\exp \left[ i\left( \eta \left( x\right) -i\gamma
_{5}\phi \left( x\right) \right) \right] \chi \left( x\right) ,~\bar{\psi}%
\left( x\right) =\bar{\chi}\left( x\right) \exp \left[ -i\left( \eta \left(
x\right) +i\gamma _{5}\phi \left( x\right) \right) \right] ,  \label{eq 1.5}
\end{equation}%
which correspond, in the path-integral framework, to the bosonization
realization in the operator approach. Under the change of variables, $\left( %
\ref{eq 1.2}\right) $ and $\left( \ref{eq 1.5}\right) $, and identity $%
\left( \ref{eq 1.4}\right) $, the Lagrangian $\left( \ref{eq 1}\right) $ is
written as%
\begin{equation*}
\mathcal{L}_{eff}\mathcal{=}\bar{\chi}i\widehat{\partial }\chi -m\bar{\chi}%
e^{-2i\gamma _{5}\phi }\chi +\frac{1}{2g}\left( \tilde{\partial}_{\mu }\phi
\right) ^{2}+\frac{1}{2g}\left( \frac{g}{\xi }\square +1\right) ^{2}\left(
\partial _{\mu }\eta \right) ^{2}+\frac{1}{2e^{2}}\phi \square ^{2}\phi ;
\end{equation*}%
however, we still have a remaining $\phi$ and fermionic fields interaction; but such interaction can be solved by an
appropriated pertubative mass expansion. However, before perform the mass expansion, we need to pay attention to some important
considerations regarding the bosonization in the path-integral framework. In such framework, it is the measure of the
original transition amplitude
\begin{equation*}
Z=N\int D\theta DA_{\mu }D\bar{\psi}D\psi \exp \left[ i\int d^{2}z\mathcal{L}%
\right] ,
\end{equation*}%
one of the most important objects in the theory and which, under the transformations $\left( \ref{eq 1.2}\right) $ and $\left( %
\ref{eq 1.5}\right) $, changes its form as:%
\begin{equation*}
DA_{\mu }=J_{A}D\eta D\phi ;D\bar{\psi}D\psi =J_{F}D\bar{\chi}D\chi ,
\end{equation*}%
with $J_{A}=\exp \left( -\bigtriangledown ^{2}\right) $, which can be
absorved into normalization constant $N$; but, the fermionic Jacobian $J_{F}$%
, has a nontrivial expression due to the axial anomaly. However, due the
abelian character of the MGTM, its evaluation is direct
using the well-known results of fermionic determinant \cite{18}; and it results into:%
\begin{equation}
J_{F}=\exp \left[ -i\frac{1}{2\pi }\int d^{2}x\left( \partial _{\mu }\phi
\right) \left( \partial ^{\mu }\phi \right) \right]   \label{eq 1.7a}
\end{equation}%
At this point, we have seen that the $\theta $-field has not participated in the bosonization process, i.e., does not have a
variables changes involving it; however, this is only possible due to our gauge choice $\left( \ref{eq 1.3}\right) $,
which leads to the identity $\left( \ref{eq 1.4}\right) $. From now on, we absorb the $\theta $-field
integration into the normalization constant. Therefore, after all changes of variables and manipulations, we then have the following transition amplitude:%
\begin{eqnarray}
Z_{MGTM}=N^{\prime }\int D\bar{\chi}D\chi D\phi D\eta \exp \Big[i\int d^{2}x%
\Big(\bar{\chi}i\widehat{\partial }\chi -m\bar{\chi}e^{-2i\gamma _{5}\phi
}\chi -\frac{1}{2g}\left( 1+\frac{g}{\xi }\square \right) ^{2}\eta \square
\eta && \notag \\
+\frac{1}{2e^{2}}\phi \square \left[ \square +M^{2}\right] \phi \Big)%
\Big],&&  \label{eq 1.8}
\end{eqnarray}%
where we have $M^{2}=e^{2}\left( \frac{1}{\pi }+\frac{1}{g}\right) $. As we can see in the expression for $Z$ $\left( \ref{eq 1.8}\right) $, the
scalar field $\eta $ is totally decoupled from the other fields; and, hence, also can be absorved in the normalization constant. Although the next step
in the path-integral quantization it is add sources to the fields on $\left(\ref{eq 1.8}\right) $, and defines the generating functional; it is sufficient to our intend -- solve exactly the bosonized MGTM and, thus, study its quantum isomorphisms -- perform a pertubative expansion in the fermionic mass, to then, finally, evaluate the respective Wightman's functions; hence, rewriting the amplitude transition $\left( \ref{eq 1.8}\right) $ in the form,
\begin{eqnarray}
Z_{MGTM} &=&N^{\prime \prime }\int D\bar{\chi}D\chi D\phi \exp \left[ i\int
d^{2}x\left( \bar{\chi}i\widehat{\partial }\chi +\frac{1}{2e^{2}}\phi
\square \left[ \square +M^{2}\right] \phi \right) \right]   \notag \\
&&\times \underset{k=0}{\overset{\infty }{\sum }}\frac{\left( -im\right) ^{k}%
}{k!}\underset{i=1}{\overset{k}{\prod }}\int d^{2}x_{i}\bar{\chi}\left(
x_{i}\right) e^{-2i\gamma _{5}\phi \left( x_{i}\right) }\chi \left(
x_{i}\right) ,  \label{eq 1.9}
\end{eqnarray}%
whose expression, immediatly, reads as:%
\begin{equation}
Z_{MGTM}=\underset{k=0}{\overset{\infty }{\sum }}\frac{\left( -im\right) ^{k}%
}{k!}\left\langle \underset{i=1}{\overset{k}{\prod }}\int dx_{i}\bar{\chi}%
\left( x_{i}\right) e^{-2i\gamma _{5}\phi \left( x_{i}\right) }\chi \left(
x_{i}\right) \right\rangle _{0},  \label{eq 1.11}
\end{equation}%
we are closer of finding the exact solution for the transition amplitude of MGTM. Here $\left\langle {\quad }\right\rangle _{0}$ stands for the vacuum
expectation value (vev)\ of an operator in a system of massless free fermions and massive free scalars. In order to evaluate $\left( \ref{eq 1.11}%
\right) $ we need to separate the bosonic and fermionic fields on the argument of vev. For this, we write
\begin{equation}
\bar{\chi}e^{-2i\gamma _{5}\phi }\chi =e^{-2i\phi }\bar{\chi}\frac{\left(
1+\gamma _{5}\right) }{2}\chi +e^{2i\phi }\bar{\chi}\frac{\left( 1-\gamma
_{5}\right) }{2}\chi ,  \label{eq 1.19}
\end{equation}%
such property leads to the new expression:
\begin{eqnarray}
Z_{MGTM} &=&\underset{n=0}{\overset{\infty }{\sum }}\frac{\left( -im\right)
^{2n}}{\left( n!\right) ^{2}}\int \left( \underset{k=1}{\overset{n}{\prod }}%
d^{2}x_{k}d^{2}y_{k}\right) \left\langle \exp \left( -2i\underset{j}{\sum }%
\left( \phi \left( x_{j}\right) -\phi \left( y_{j}\right) \right) \right)
\right\rangle _{0}^{bos}\times   \notag \\
&&\times \left\langle \underset{i=1}{\overset{n}{\prod }}\bar{\chi}\left(
x_{i}\right) \frac{\left( 1+\gamma _{5}\right) }{2}\chi \left( x_{i}\right)
\bar{\chi}\left( y_{i}\right) \frac{\left( 1-\gamma _{5}\right) }{2}\chi
\left( y_{i}\right) \right\rangle _{0}^{ferm}.  \label{eq 1.12}
\end{eqnarray}%
The simplest term to evaluate $\left( \ref{eq 1.12}\right) $, is the
fermionic contribution, where the fermionic Wightman function is simply,
the free fermion propagator%
\begin{equation}
S_{F}\left( x\right) =-\frac{1}{2\pi }\frac{\gamma ^{\mu }x_{\mu }}{x^{2}};
\label{eq 1.20}
\end{equation}%
to evaluate the fermionic part, we decompose the spinors in their components,
as:%
\begin{eqnarray}
\bar{\chi}\frac{\left( 1+\gamma _{5}\right) }{2}\chi  &=&\bar{\chi}_{1}\chi
_{1},  \label{eq 1.21a} \\
\bar{\chi}\frac{\left( 1-\gamma _{5}\right) }{2}\chi  &=&\bar{\chi}_{2}\chi
_{2},  \label{eq 1.21b}
\end{eqnarray}%
such manipulation leads to the well-known result:
\begin{equation}
\left\langle \underset{i=1}{\overset{n}{\prod }}\bar{\chi}_{1}\left(
x_{i}\right) \chi _{1}\left( x_{i}\right) \bar{\chi}_{2}\left( y_{i}\right)
\chi _{2}\left( y_{i}\right) \right\rangle _{0}^{ferm}=\frac{1}{\left( 2\pi
i\right) ^{2n}}\frac{\overset{n}{\underset{i>j}{\prod }}\left(
c^{2}\left\vert x_{i}-x_{j}\right\vert ^{2}\left\vert y_{i}-y_{j}\right\vert
^{2}\right) }{\underset{i,j}{\overset{n}{\prod }}\left( c\left\vert
x_{i}-y_{j}\right\vert ^{2}\right) };  \label{eq 1.15}
\end{equation}%
where $c=e^{-\gamma }$, with $\gamma $ the Euler-Mascheroni constant. Now,
to calculate the bosonic part, we must first evaluate the scalar Wightman
function; defining it as:%
\begin{equation*}
\frac{1}{e^{2}}\square \left[ \square +M^{2}\right] \Delta \left( x\right)
=-\delta ^{\left( 2\right) }\left( x\right) ,
\end{equation*}%
see equation $\left( \ref{eq 1.9}\right) $; its solution is ready obtained,%
\begin{equation*}
\Delta \left( x\right) =\lambda ^{2}\int \frac{d^{2}k}{\left( 2\pi \right)
^{2}}\left[ \frac{1}{k^{2}}-\frac{1}{k^{2}-M^{2}}\right] e^{-ikx}=\lambda
^{2}\left( \Delta \left( 0;x\right) -\Delta \left( M^{2};x\right) \right) ,
\end{equation*}%
\begin{equation}
\Delta \left( x\right) =\lambda ^{2}\left( -\frac{1}{4\pi }\ln \left(
M^{2}c^{2}x^{2}\right) -\frac{1}{2\pi }K_{0}\left( \sqrt{M^{2}x^{2}}\right)
\right) .  \label{eq 1.7}
\end{equation}
where $\lambda ^{2}=\frac{\pi }{1+\frac{\pi }{g}}$, and $K_{0}\left(
z\right) $ is the second-class modified Bessel's function \cite{25}. Then, we obtain
the well-known result,
\begin{equation}
\left\langle \exp \left( -2i\underset{j}{\sum }\left( \phi \left(
x_{j}\right) -\phi \left( y_{j}\right) \right) \right) \right\rangle
_{0}^{bos}=\exp \left[ 4\underset{i>j}{\sum }\left( \Delta \left(
x_{i}-x_{j}\right) +\Delta \left( y_{i}-y_{j}\right) -\Delta \left(
x_{i}-y_{j}\right) \right) \right] ,  \label{eq 1.13}
\end{equation}%
and substituing the $\Delta \left( x\right) $ expression, Eq.$\left( \ref{eq 1.7}\right) $,
into the above equation, yields to the following expression,%
\begin{eqnarray*}
\left\langle \exp \left( -2i\underset{j}{\sum }\left( \phi \left(
x_{j}\right) -\phi \left( y_{j}\right) \right) \right) \right\rangle
_{0}^{bos} =
\end{eqnarray*}
\begin{eqnarray}
&=&\left[ Mc\right] ^{\frac{2n\lambda ^{2}}{\pi }}\frac{\overset{n%
}{\underset{i>j}{\prod }}\left\vert x_{i}-x_{j}\right\vert ^{-\frac{2\lambda
^{2}}{\pi }}\left\vert y_{i}-y_{j}\right\vert ^{-\frac{2\lambda ^{2}}{\pi }}%
}{\underset{i,j}{\overset{n}{\prod }}\left\vert x_{i}-y_{j}\right\vert ^{-%
\frac{2\lambda ^{2}}{\pi }}}  \label{eq 1.14a} \\
&&\times \exp \left[ -\frac{2\lambda ^{2}}{\pi }\underset{i>j}{\sum }\left(
K_{0}\left( M;\left\vert x_{i}-x_{j}\right\vert \right) +K_{0}\left(
M;\left\vert y_{i}-y_{j}\right\vert \right) -K_{0}\left( M;\left\vert
x_{i}-y_{j}\right\vert \right) \right) \right] .  \notag
\end{eqnarray}%
Therefore, with the results $\left( \ref{eq 1.15}\right) $ and $\left( \ref%
{eq 1.14a}\right) $, we finally get the resulting expression for the
transition amplitude $\left( \ref{eq 1.12}\right) $, which is written as:
\begin{eqnarray}
Z_{MGTM} &=&\underset{n=0}{\overset{\infty }{\sum }}\frac{1}{\left(
n!\right) ^{2}}\left( \frac{m}{2\pi c}\left[ Mc\right] ^{\frac{\lambda ^{2}}{%
\pi }}\right) ^{2n}\int \left( \underset{k=1}{\overset{n}{\prod }}%
d^{2}x_{k}d^{2}y_{k}\right) \frac{\overset{n}{\underset{i>j}{\prod }}%
\left\vert x_{i}-x_{j}\right\vert ^{2\left( 1-\frac{\lambda ^{2}}{\pi }%
\right) }\left\vert y_{i}-y_{j}\right\vert ^{2\left( 1-\frac{\lambda ^{2}}{%
\pi }\right) }}{\underset{i,j}{\overset{n}{\prod }}\left\vert
x_{i}-y_{j}\right\vert ^{2\left( 1-\frac{\lambda ^{2}}{\pi }\right) }}
\notag \\
&&\times \exp \left[ -\frac{2\lambda ^{2}}{\pi }\underset{i>j}{\sum }\left(
K_{0}\left( M;\left\vert x_{i}-x_{j}\right\vert \right) +K_{0}\left(
M;\left\vert y_{i}-y_{j}\right\vert \right) -K_{0}\left( M;\left\vert
x_{i}-y_{j}\right\vert \right) \right) \right] .  \label{eq 1.16}
\end{eqnarray}%

\section{Isomorphisms of Bosonized MGTM}
\label{sec:2}

At this point, we can study the equivalence of MGTM by applying the following limits: $g\rightarrow\infty$ and $e^{2}\rightarrow \infty$, to reproduce
the massive Schwinger and Thirring models, respectively, into the bosonized transition amplitude of MGTM $\left( \ref{eq 1.16}\right) $.
But first, we have that the function $K_{0}\left( z\right) $ has the following asymptotic limits \cite{25}:%
\begin{eqnarray*}
K_{0}\left( z\right)  &\rightarrow &-\ln \left( z\right) -\gamma
;~~z\rightarrow 0, \\
K_{0}\left( z\right)  &\rightarrow &0;~~z\rightarrow \infty .
\end{eqnarray*}
Now, by the limit: $e^{2}\rightarrow \infty \leftrightarrow M^{2}\rightarrow \infty $, was proved that GTM reproduces the Thirring model at
classical and quantum levels; here, from applying the limit into equation $\left( \ref{eq 1.16}\right) $, we get:%
\begin{eqnarray}
Z_{MTM} &=&\underset{e^{2}\rightarrow \infty }{\lim }Z_{MGTM}=\underset{n=0}{%
\overset{\infty }{\sum }}\frac{1}{\left( n!\right) ^{2}}\left( \frac{m}{2\pi
}\right) ^{2n}\int \left( \underset{k=1}{\overset{n}{\prod }}%
d^{2}x_{k}d^{2}y_{k}\right)   \label{eq 1.17} \\
&&\times \frac{\overset{n}{\underset{i>j}{\prod }}c^{2\left( 1-\frac{\lambda
^{2}}{\pi }\right) }\left\vert x_{i}-x_{j}\right\vert ^{2\left( 1-\frac{%
\lambda ^{2}}{\pi }\right) }\left\vert y_{i}-y_{j}\right\vert ^{2\left( 1-%
\frac{\lambda ^{2}}{\pi }\right) }}{\underset{i,j}{\overset{n}{\prod }}c^{1-%
\frac{\lambda ^{2}}{\pi }}\left\vert x_{i}-y_{j}\right\vert ^{2\left( 1-%
\frac{\lambda ^{2}}{\pi }\right) }},  \notag
\end{eqnarray}%
which reproduces the well-known result of the bosonized massive Thirring model \cite{6,20}. Now, for the limit: $g\rightarrow \infty $, which reproduces
the Schwinger model in the classical and quantum levels, yields to:%
\begin{eqnarray}
Z_{MSM} &=&\underset{g\rightarrow \infty }{\lim }Z_{MGTM}=\underset{n=0}{%
\overset{\infty }{\sum }}\frac{1}{\left( n!\right) ^{2}}\left( \frac{m\mu }{%
2\pi }\right) ^{2n}\int \left( \underset{k=1}{\overset{n}{\prod }}%
d^{2}x_{k}d^{2}y_{k}\right)   \label{eq 1.18} \\
&&\times \exp \left[ -2\underset{i>j}{\sum }\left( K_{0}\left( \mu
;\left\vert x_{i}-x_{j}\right\vert \right) +K_{0}\left( \mu ;\left\vert
y_{i}-y_{j}\right\vert \right) -K_{0}\left( \mu ;\left\vert
x_{i}-y_{j}\right\vert \right) \right) \right] ,  \notag
\end{eqnarray}%
where $\mu =e^{2}/\pi $; and, then, we also obtain the well-known result of the bosonized massive Schwinger
model \cite{2,20}.

With the results, Eqs.$(\ref{eq 1.17})$ and $(\ref{eq 1.18})$, we have proved the isomorphism between the MGTM
and the massive Thirring and Schwinger models, respectively; furthermore, we have extended the previous results
regarding the isomorphisms of the massless GTM \cite{14}, to the massive one.

\section{Remarks and conclusions}
\label{sec:3}

A path-integral bosonization of the massive gauged Thirring model was presented. Following a well-known, however well-established, script of path-integral
bosonization: separation of the gauge field in its divergence and divergenceless parts, and its partial-separation with the fermions,
through appropriated change of variables into fermions; such fermionic change of variables induces a chiral Jacobian in the measure of transition
amplitude; choice of a gauge condition, in the MGTM case, the $R_{\xi}$-gauge -- which allowed that the $\theta$-field was decoupled from other fields. Afterwards, we have made a pertubative expansion in the fermionic mass in the transition amplitude, allowing, then, that the resulting vacuum expectation
value of operators were easily evaluated through the well-known Wightman's functions. Furthermore, we performed in the final transition amplitude expression, equation $\left( \ref{eq 1.16} \right) $, an analysis regarding the strong coupling regime of the model; which, after the appropriated limits study, resulted the bosonized massive Schwinger model, for $g\rightarrow \infty $, and the bosonized massive Thirring model, for $e^{2}\rightarrow \infty $.
With these results, we generalized the previous results regarding the isomorphisms of massless GTM \cite{14} to the massive case.
Another nonpertubative properties involving two-dimensional massive fermionic field theories are under analysis; in particular, the extension
of previous analysis of MGTM to the case of $T \neq 0$, at both Green's functions and bosonization scenarios. We also believe that the extension of bosonization of MGTM, presented here, to the non-abelian case deserves a careful treatment \cite{17}. Progress regarding these issues are under development, and will be reported elsewhere.

\subsection*{Acknowledgements}

The authors would like to thank Professor Rodolfo Casana for carefully reading the manuscript. R.B. thanks CNPq for full support and B.M.P. thanks CNPq and CAPES for partial support.

\end{document}